# Electrification of Commercial E-buses by Utilizing Stationary Battery Energy Storage Systems for Mass Transportation Network in Los Angeles

Amirhossein Ahmadian [1,*], Atharva Sunil Deo[1], Paawan Garg[1], Omar Elayyan[1]

[1] Department of Mechanical and Aerospace Engineering, University of California, Los Angeles, CA, 90095, USA
*Corresponding Author: aahmadian@ucla.edu

## Abstract

The 2028 LA Olympics is approaching, and the mass transportation network requires to be expanded. The green electric bus transportation system is gaining increasing attention as an essential step to mitigate emission concerns. However, the introduction of a large electric bus fleet will increase the peak load demand, which will adversely affect the grid. This research addresses the issue of an increase in the peak load demand due to the introduction of the electric bus fleet. Therefore, the potential implementation of stationary battery energy storage systems is investigated to shave the peak load demand by providing energy during peak hours.

Moreover, a fleet management software implementation has been demonstrated for a test case scenario in Los Angeles. As a result, based on specific energy demand values, battery hardware options, standard e-bus driving cycles and profiles, and local bus route characteristics, the optimal battery specifications, and power rates have been addressed. Moreover, techno-economic analysis has been carried out to demonstrate the cost-benefit analysis.

# Introduction

## Motivation

The 2028 Olympic games are just eight years way, and athletes are gearing up for the most significant sports event. Therefore, before athletes, Los Angeles needs to gear up itself. L.A. 2028 will not be likely without the accomplishment of the expansion of Los Angeles County's mass transportation network and the planned expansion [1]. Also, E-mobility is currently seen as the most promising technology to reduce exhaust emissions in transportation. Electric buses are replacing conventional diesel-fueled buses at an accelerating rate that is outpacing the adoption of battery-powered cars. Electric bus chargers and charging stations market is rapidly growing, which is Expected to Reach $12.3 Billion by 2025 and increase at a CAGR of 9.7% during the forecast period published by P&S Intelligence. The growth of the market is chiefly propelled by ample government initiatives and regulations, immediate implementation of electric buses in public fleet. Moreover, numerous investments toward the development of electric bus charging stations across the world [2].

There is a number of factors behind the potentially rapid growth of the e-bus charging stations market. They comprise government initiatives and regulations, rapid adoption of electric buses in the public feet, and several investments. Presently, the e-bus charging market is at the early stage of advancements, and it substantially relies on government funding and other beneficial schemes [3]. In this study, research on the electrification of the public bus network in Los Angeles will be carried out.

## Problem Definition

This research project aims to address the impact of the electric bus (e-bus) charging infrastructure level of service and costs and how could this be regulated efficiently. Charging infrastructure is the bottleneck for electric bus mass deployment. Therefore, this study attempts to provide insights into the choice of charging infrastructure, method, and battery specifications. In this regard, an increase in the peak load demand and its negative impact on the grid due to en-route charging are discussed in particular. To address the aforementioned issues, we delve in to charging e-buses through en-route chargers by using Stationary Battery Energy Storage Systems (BESS).

Also, based on the charging mechanisms and predefined bus routes, and electrification distributions (electric city and/or regional buses, electric BRT vehicles), a software implementation of an optimized scheduling framework and allocated charging stations locations with optimized battery specifications for electric buses have to be achieved.

# Research Proposal

## Software Solutions

A new fleet management software should be developed. It will take different scenarios of refueling e-buses and charging mechanisms into account. Moreover, based on specific energy demand values, battery hardware options, standard e-bus driving cycles, and profiles, and local bus route characteristics, the optimal battery specifications, and power rates will be populated. This software then could be integrated into BLAST-BTM software, developed by National Renewable Energy Laboratory, and will be customized to the interest of Greater Los Angeles. Also, battery performance modeling will be studied over a long period by considering thermal and degradation models and focusing on the electrical performance of the battery. This software investigates the use of Li-ion battery degradation and longevity in community energy storage (CES) applications.

In general inputs to the proposed software are divided into four segments:

## Hardware Options

This feature lets the user define the cost associated with the system installation and maintenance and the quantitative features of the battery system. The cost includes the upfront installation cost, and the incentives gained and the maintenance of the system. The hardware configuration is defined by the inverter efficiency and the size of the energy storage system (ESS) by defining the energy fraction and the duration for full discharge .

## Route Characteristics

There are various parameters that need to be taken into account for the electrification of any commercial vehicle or any public transit vehicle, like the operating range, schedules, flexibility, charging time, cost, grid capacity, fleet size, charging interfaces, and passengers. Here, the ultimate goal is to make the e-buses run productively and efficiently.

In order to determine which type of charging method is practical for each application, an evaluation of a work-day duty cycle and available charge times should be performed. This feature will let the user enter standard e-bus driving cycles and profiles and local bus route characteristics. A brief overview of the types of **charging scenarios and characteristics** that are considered in our project are mentioned below:

## En-route Charging

Demand for long-running e-buses has given rise to the en-route charging, or so-called opportunity charging, of batteries. En-route charging occurs when the bus stops momentarily to pick up and drop off passengers. In these situations, e-buses require higher power to deliver charge in such a short time. So far, all en-route chargers utilize DC chargers and operate between 150KW and 600KW [4]. En-route charging infrastructures are designed to meet a short driving range. Thus, batteries need to be sized in order to store enough energy to reach the next charging point. En-route opportunity charging can be done in several different ways: conventional charger located at terminals or on-route, magnetic induction, catenary, or overhead charging systems at bus stops. Usually, city buses operate daily with fixed routes in downtown areas. Ultra-fast charging those buses in terminals during driver breaks and between shifts provides multiple charging opportunities throughout the day without impacting operations [5].

As stated previously, there are multiple ways to fast charge the batteries; all of them are aimed at expediting the charging process, reducing downtime compared to conventional charging. Opportunity charging involves charging the batteries at 150kW – 600kW charge rate whenever possible. Due to frequent charging and the intention to limit battery gas generation, opportunity chargers are typically set to charge the batteries up to 80% - 85% SOC throughout the day and back to 100% once a day (e.g., during night hours) [4] .

Opportunity charging is the right choice for extending shift operations where battery changing can be eliminated. In addition, opportunity charging extends the run times of aging batteries and recoups the lost capacity that comes with age.

One of the widely used opportunity charging methods is called overhead charging. As its name suggests, it is to charge the e-buses using pantographs in high charging rates. This method is adopted by many public transit systems around the world, such as New Flyer New York, TOSA in Geneva, and the city bus of the Swedish city Gothenburg. Also, LA Metro bus (Figure 1) is planning to adopt automated SAE3105-1 en-route charging operations to optimize time to charge
First, the overhead charging system is designed to be a part of regular bus stops; thus, the charging infrastructure has minimal impact on the original bus stop design.
Second, the fixed route of city buses can effectively reduce the amount of charging stations needed. Setting charging pantographs on each bus stop is a doable way; however, with the high- power supply infrastructures, it is possible to set up charging stations only at the start and the final stations [6] (Figure 2, Figure 3).

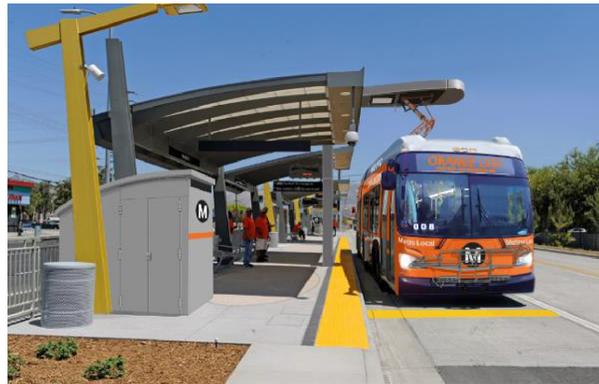

*Figure 1- Orange line electric bus equipped with SAE3105-1 charging station [6].*

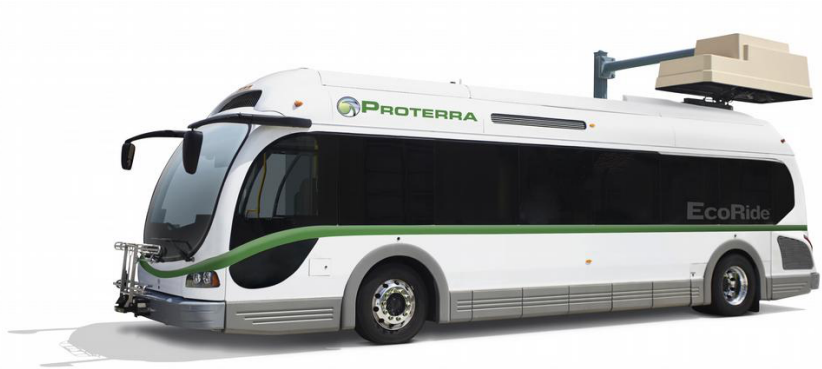

*Figure 2- Proterra en-route charging station, compliant with J3105 [7].*

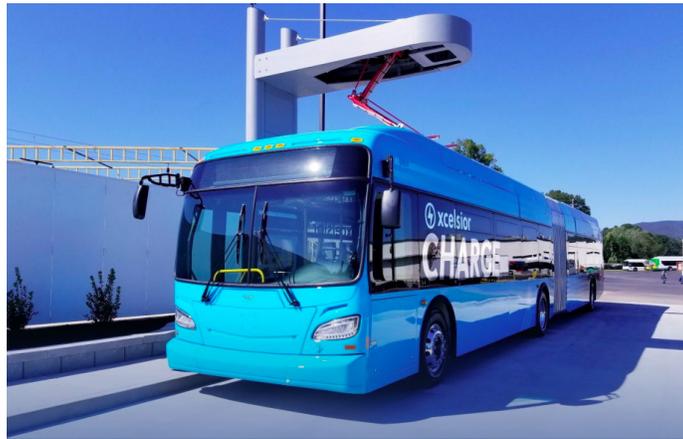

*Figure 3- New Flyer Rapid catenary configuration for en-route bus chargers, compliant with OppCharge, and SAE charging protocols [8].*

Depot Charging

Depot charging is mostly concerned during the night time when the batteries have enough time to charge for 8-10 hours, work for a shift of 8-10 hours, and another 8-10 hours of rest and cooling time. By doing so, the chargers will be enabled to pump power into the batteries to a 100% state of charge (SOC) daily [9].

This kind of charging method fits the one-shift application public transit such as the school bus since they only run in the morning and the afternoon and have an average mile of 60-70 miles per day. If multi-shift is needed, users may need one or more battery modules per bus.

Although the construction cost is way lower compared to fast charging systems, the electrification of this type of operating mode still can be costly because the large amount sits parked time of one-shift application transit. This increases the timeline to earn back the cost difference through fuel and maintenance savings. Nevertheless, one way the operator can make electric buses more affordable is through power sales to electric utilities when the e-buses are not in use.

It usually happens when buses require to be charged almost to a 100% state of charge (SOC) overnight. It usually takes an average of 8-10 hours to fully charge the battery. In this scenario, the battery is charged over a 5 to10 hour period, rests for another 5-10 hours, and is used over a 10-hour shift. As such, conventional charging is ideal for one-shift applications where no battery charging/ changing is required. Also, in multi-shift operations (2 or 3 shifts) user needs to charge more than once for the bus or requires battery changing between shifts [10]. In the following figure, some examples of depot charging battery storage systems while fueling buses are brought (Figure 4, Figure 5).

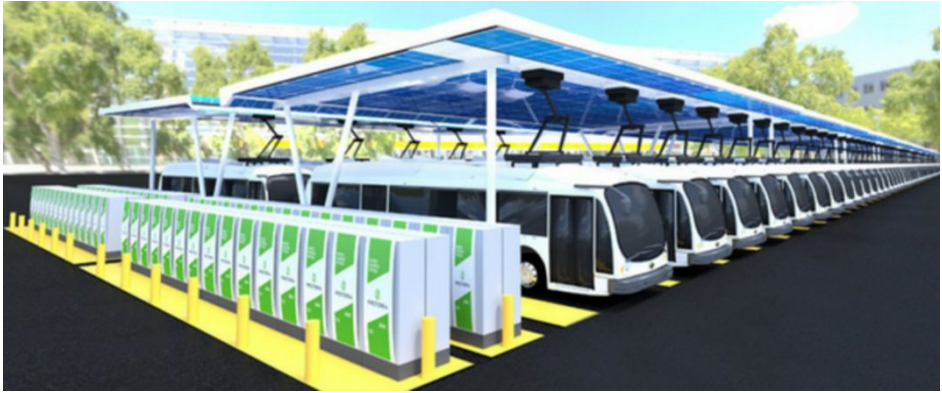

*Figure 4- Proterra depot charging configuration, compliant with J3105, including PV cells and stationary battery storage* [7].

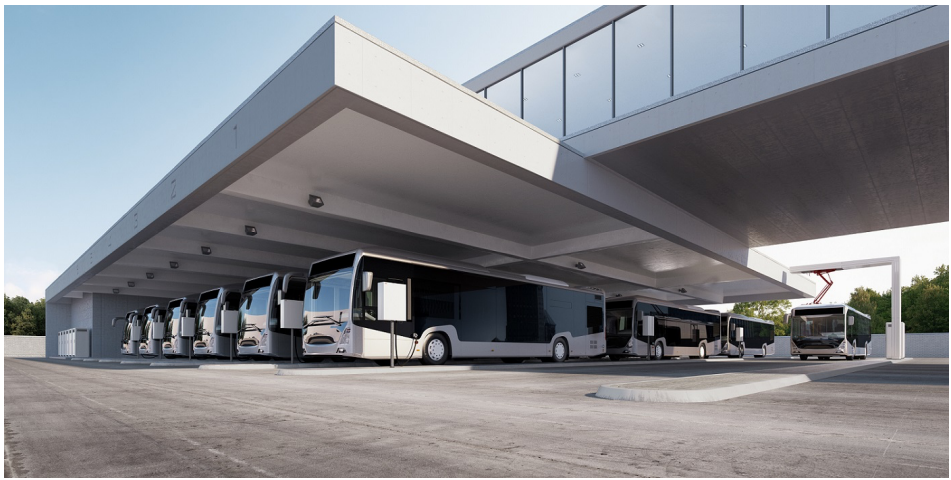

*Figure 5- ABB depot charging station, purposed for overnight charging* [11].

Station Characteristics

This input parameter serves as either an alternative to route characteristics or a concurrent dataset to provide higher output veracity. For each input set, one of two types of stations can be chosen, designated either as en-route stations or depots, and chosen by entering a specific location or set of coordinates to be reverse geocoded into a location. For each station, various parameters, comprising bus arrival and departure times, bus make and model (for ascertaining charging rate and comparable capabilities), average passenger traffic as a function of time of day, as well as the time of year to accommodate seasonal traffic changes can be inputted.

In contrast to the utilization of standard e-bus driving cycles in the context of localized transportation, station characteristics provide for comprehensive stationary energy storage systems for a given station on a route. Thereby, the tradeoff concerns an increase in result veracity regarding a particular station but disjoint results for the stations on a given route.

### Demand Profile and Rate Structure

This lets the user define the power demand per interval for a year. Alternatively, the tool can pull data from the EnerNOC database that contains historical demand data spanning 98 load sites. This input will be specified by entering a specific location. Also, Southern California Edison pricing has been used for the sake of techno-economic analysis, which will be elaborated in the later sections.

Currently, all commercial, industrial, and agricultural customers in California are mandated to be compliant with the time-of-use plan. Also, residential customers can choose to be on to the time of use plans by contacting their utility. Battery storage systems help us to shift energy usage from peak hours to off-peak hours, which could reduce the energy bill by switching to a time-of-use rate plan.

### Software GUI Design

The formerly discussed aspects of reducing the customer's utility cost comprise the primary aspects of the proposed software design to meet the customer's goal. This tool will likely be a cloud-based application to enable straightforward access for employees and provide convenience. In general, discussion of software design can be divided into the following:

### UI/UX Design

For the aforementioned parameters, the user interface will consist of various input fields and visualizations.

Hardware option input needs will be met with editable text fields for budget and price range, as the first step in result generation.

Route characteristics input can be in the format of .txt or .json files, in addition to application-based input tools, comprising text fields for station and depot stops on a given route, and desired time of arrival at each instance. Additional information that can be inputted includes the average time of arrival in each case and passenger traffic between stops. Station stop information will be concurrently displayed on a map, utilizing the Google Maps API, as the information is entered, and various metrics, including route load and time, will be plotted on graphs.

Station characteristics input can be supplemental to route characteristics or serve as the primary means of output determination. Text Fields will be provided for station location information or inputted through .txt or .json files, and this information will be displayed on a map. For each input station, a drop-down menu allowing for the selection of a type of station, en route or depot, will appear, and parameters, comprising bus arrival and departure times, bus make and model, and average passenger traffic at different times of day can be entered. Information including passenger traffic and comparable metrics will be plotted on a graph as a function of time of day.

The field for demand profile will consist of a location-entry text field, which will pull data from the EnerNOC database in this regard and display it.

The input section for rate structure will allow for facility utility costs to be entered as well.

Input/Output

As previously defined, input parameters comprise hardware options, inclusive of budget and price range constraints, route characteristics, station characteristics, generalized demand profile, and aspects of rate structure and facility utility cost. Such input parameters would concern various UI elements to enable ease of use and convenience in general.

The output parameters of the tool comprise a generalized ESS solution. This solution would comprise stationary storage products, including manufacturing and model details, inverter characteristics for a particular station, and charging infrastructure, if not already specified by the provider of e-buses. In the following figure (Figure 6), a quick review of the tool can be viewed.

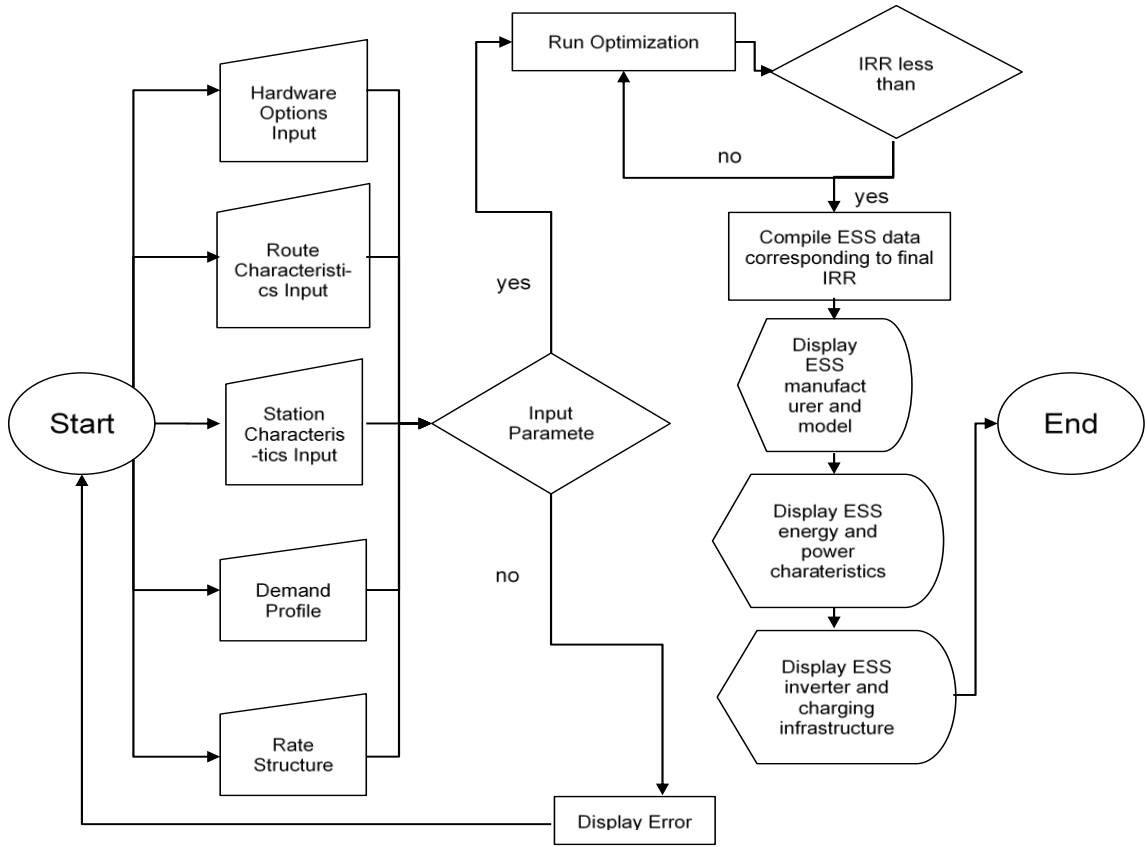

*Figure 6- Flowchart with a quick view of how the tool will function.*

The proposed software tool should provide the optimal energy and power points at which the rate of return is maximum. The optimization takes the following steps, and in general, outputs to this software considers the following:

1. Lowest energy, highest power to energy ratio ESS is considered, to begin with
2. Seeks the optimal power-to-energy ratio by incrementally decreasing the ratio while holding the ESS energy level constant
3. The algorithm assumes that IRR will monotonically increase to a maximum value as the ratio is decreased, then decrease thereafter, it, therefore, stops whenever the IRR is found to be less than the previous one
4. The maximum IRR for this ESS energy level is kept in memory
5. The above steps are repeated for the next energy level
6. If the highest IRR for the current energy level is less than that of the last one, the algorithm terminates, and the highest IRR is provided as the result

The IRR should not only account for the savings made by peak shaving that results in reduced utility bills but also needs to consider E-bus annual income. This must be included in the mathematical calculation followed by the following formula:

$$Upfront\ Cost - Incentives = \sum_{n=1}^{N} \frac{Annual\ Savings + Annual\ Net\ Income}{(1 + IRR)^n}$$

Data Flow Analysis

Some of the parameters that are raised and required for the optimization problem are listed below:

1- Energy usage (KWh) for the e-Bus operation between 8:00 a.m. and 10:00 p.m.
2- The number of batteries required to sustain the charging system during peak loads.
3- Monthly saving in energy cost from power arbitrage using Energy Storage.
4- Anticipated battery warranty lifecycle-based upon the charge/discharge process.
5- ROI (Return of Investment) to reach the breakeven.
6- Energy Storage lifetime cost/benefit.
7- Energy cost vs. time in a 24-hour time-frame

Out of these parameters, the state of charge and location are considered as the real-time data, since the data varies continuously in this case. But, for the other parameters, the data is considered fixed (constant), since the data is computed after taking averages to account for most of the variations.

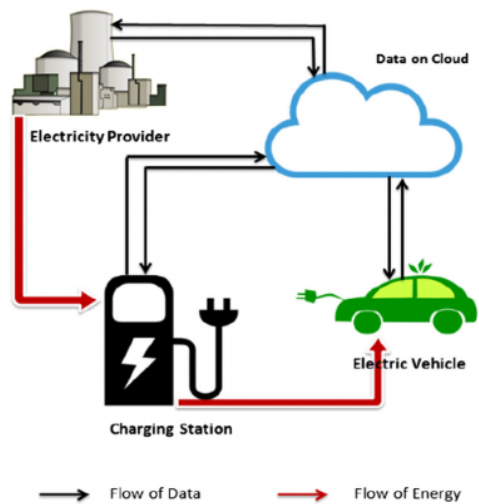

*Figure 7- Data and Energy Flow in a cloud-based management platform [12].*

A communication network is required to ensure the systematic scheduling of the buses. One way of communication is through the use of a cloud platform. The data can be generated in real-time on the bus using a controller and DAQ (Data Acquisition System), which provides information about the location and state of charge. On the stationary battery system, the state of charge can be monitored in a similar way. This information can be centrally stored in a cloud database for all the buses using the same route. Then this information can be retrieved by a computer system to implement the optimization algorithm taking the real-time data as the input parameters. A sliding window approach can be implemented, which maintains the size of the dataset by continuously removing the old data and including the new data [12]. The output of the algorithm will be an optimized solution for the battery size. Therefore, the battery capacity can be chosen after considering a factor of safety to take any kind of unprecedented variations into account.

## BESS Control System

The primary facets of an effective battery management system (BMS), in the contexts of power distribution quality, battery lifespan, and cell balancing, comprises hardware aspects that can measure fundamental parameters such as cell voltage, current, and temperature and software aspects that can record and interpret this data and perform sophisticated tasks. Such a system would require efficient interplay between advanced sensory systems and software-initialized reactions to scenarios, providing for active control of passive devices such as electrochemical batteries. This section in the future final report will delve into the existing architectures for hardware and software in terms of battery system controls and provides an overview of presently proposed and implemented designs [13].

## General Communication Requisites

Various communication protocols may be employed for effective data transmission between sensors and supervisory microcontrollers. Various forms of serial communication, such as RS232, can be used, as well as CAN bus and Modbus communications [14].

## General BMS Software Requisites

Given the aforementioned discussions on the hardware and sensory aspects of an effective BMS, the next point of the analysis is that of capable controls software. General BMS controls software requires the following [13, 15-17]:

1. Cell Balancing Capabilities
    1. Hardware-based cell balancing
    2. State of charge prediction
    3. Algorithmic cell balancing
2. Sensory data extraction
    1. General communication protocols
3. Fault handling

## General Cloud-Based Management Software Requisites

The formerly discussed aspects of reducing the customer's utility cost comprise the primary aspects of the proposed software design to meet the customer's goal. This tool will likely be a cloud-based application to enable straightforward access for employees and provide convenience.

# Result and Analysis

## End-User load profile and rate structure

According to the California Public Utilities Commission's mandatory energy storage procurement targets, its IOUs are required to collectively procure 200 MW energy storage in the customer-side domain by 2020 [18].

This research project proposes methods to estimate the potential benefits of a battery storage system for the behind-the-meter application. Also, it determines the optimal energy and power capacity. The capability of lowering peak demand from battery provides an excellent opportunity to cut the electricity bill. Therefore, it is desired to determine the most economical battery size considering both benefits and costs.

It should be noted that the total capital cost consists of not only the investment and installation cost but also the total O&M. In general, total savings in electricity bills do not vary significantly with load patterns. Moreover, the demand charge reduction depends on load shape. In some instances, when the battery size is large enough to provide sufficient power, it can significantly lower the monthly demand.

Also, the saving in demand charge generally increases as the load factor decreases. It largely depends on seasonal variation while the demand charge is typically on a monthly basis. Moreover, it is worth to mention that rate tariff varies significantly from one geographic location to another. This impacts on change in battery economic benefits. Here the following rate is assumed for electric vehicle plan (Table 1, Table 2):

Table 1- June to September (4 months) rates per kWh according to southern California Edison [19].

| Weekdays | | | Weekends | | |
|---|---|---|---|---|---|
| 13 | 38 | 13 | 13 | 27 | 13 |
| 8 am - 4 pm | 4 pm – 9 pm | 9 pm – 8 am | 8 am – 4 pm | 4 pm – 9 pm | 9 pm – 8 am |

Table 2- October to May (8 months) rates per kWh according to southern California Edison [19].

| Weekdays | | | Weekends | | |
|---|---|---|---|---|---|
| 12 | 35 | 12 | 12 | 35 | 12 |
| 8 am - 4 pm | 4 pm – 9 pm | 9 pm – 8 am | 8 am – 4 pm | 4 pm – 9 pm | 9 pm – 8 am |

The question raised in this optimization problem raised here can be listed as the following:

1. How much energy is used between the e-bus operation time between 8 a.m. and 10 p.m. measured in kWh?
2. How many batteries are required to cover our on-peak loads?
3. What is the monthly energy cost savings from power arbitrage using energy storage?
4. What is the anticipated battery warranty lifecycle based on this charge/discharge process?
5. How long until the return on investment (ROI) reaches breakeven?
6. What is our energy storage lifetime cost/benefit?

Techno-Economic Analysis

To cut the load on the grid, en-route charging systems can be equipped with stationary battery energy storage systems (BESS), which will provide the juice to the e-buses in the daytime and will get themselves charged in the night-time. This can also compensate for the dip in the duck curve, which happens during the night-time.

The capability of lowering the peak demand through the use of a stationary battery system provides an excellent opportunity to cut the electricity bill because the rate of electricity in the night is significantly low, and thus charging the stationary battery system wouldn't become costly.

Therefore, it is desired to determine the most optimal battery size considering both its benefits and the cost.

Lithium-titanate batteries (LTO) are known for tremendous power output, extended life cycle, and reliability. These may be the ideal choice for en-route charging, due to the high-power output necessary and the fact they will be cycling several times per day, dramatically cutting the life short for most other battery options. Typically, a fully electric bus can be charged within 5-10 min through each LTO batteries [20]. The following table gives the proposed battery specifications for En-route charging.

*Table 3- Proposed battery specifications for short range purposes.*

| Vehicle Type | City e-Bus |
|---|---|
| Battery Type | LTO battery 120 KWh/250KWh |
| Mileage | 150km per day |
| Charging time | ~ 5 min (from 60% to 100% after arriving at each bus stop) |
| Operation Environment | Greater Los Angeles |
| Battery Cycle Life | > 12000 cycles battery with less than 10% capacity fade |

Let's assume the following time-scenario for charging e-buses with the above proposed battery specifications:

8:30 am 45 kWh at 500 kW

9:10 am 35 kWh at 500 kW

10:15 am 50 kWh at 500 kW

11:45 am 60 kWh at 500 kW

12:30 pm 70 kWh at 500 kW

1:15 pm 40 kWh at 500 kW

2:45 pm 40 kWh at 500 kW

4:00 pm 60 kWh at 500 kW

5:15 pm 20 kWh at 500 kW

Then the e-buses have the following power usage:

- Daytime usage: 8 a.m. – 4 p.m. = 400 kWh
- Evening usage: 4 p.m. – 9 p.m. = 20 kWh
- Nighttime usage (Battery recharge): 10 p.m. – 4 a.m. = 360 kWh
- 24-hour total usage: 780 kWh

According to SCE [19], the total bill comes to ~ **$106 per day** or an average **monthly bill of ~$3200**.

On the batteries side, we have taken arbitrage-use conditions of 600 MWh of battery throughput. Throughput is defined as the energy moved through the battery during each roundtrip discharge/charge cycle. To calculate the length of the warranty period, we divide half of 600 MWh by annual KWh discharge. Then the equations become:

- **Annual KWh Discharge**: $Daily\ Discharge \times 365 = 780\ KWh \times 365$

$$= 284{,}700\ KWh/year$$

- **Warranty Period:** $\dfrac{600 MWh/2}{284700\ KWh/year} = 1.05\ years$ (ignoring losses)

Next, if the profit/cost benefit is to be realized, the battery installation should be less than $40,625 (we're using a typical federally subsidized installation cost). If the installation costs are higher than this number, then break-even may not be brought to fruition. This can be seen from the following calculations;

- **Lifetime Discharge:** (Warranty Period * Annual KWh Discharge)

$$= (1.05 \text{ x } 284{,}700 \text{ kWh/year}) = 298935\ KWh$$

- **Cost per KWh:** (Battery Installation Cost / Lifetime Discharge)

$$= (\$40625\ /\ 298935\ KWh) = \$0.136/kWh$$

- **Daily Power Cost from Battery:** (Daily discharge rate * Cost per KWh)

$$= (780\ kWh \text{ x } \$0.136/kWh) = \$106/day$$

In summary, the total bill comes to ~ **$106 per day** or an average monthly bill of ~**$3180.** Therefore, it can be seen that for the cost benefits to be realized, and the battery installation costs are to be minimized to less than $40,625.

Another aspect of the cost analysis is the actual installation cost of the LTO batteries, which is near $600/KWh. With this actual scenario, achieving break-even might take a lot of time or at the worst, it may not be perceived. This boils down to the dichotomy of efficiency and performance.

The domain of this research is found upon the performance partition as we are advocating a stationary BESS that will help cut the load on the grid. Cost benefits, however, are still not a part of the scope of this project.

Now, in order to simplify the power extracted from the grid without implementing batteries as compared to employing battery energy storage systems; the following power amount from the battery support can be saved Figure 6:

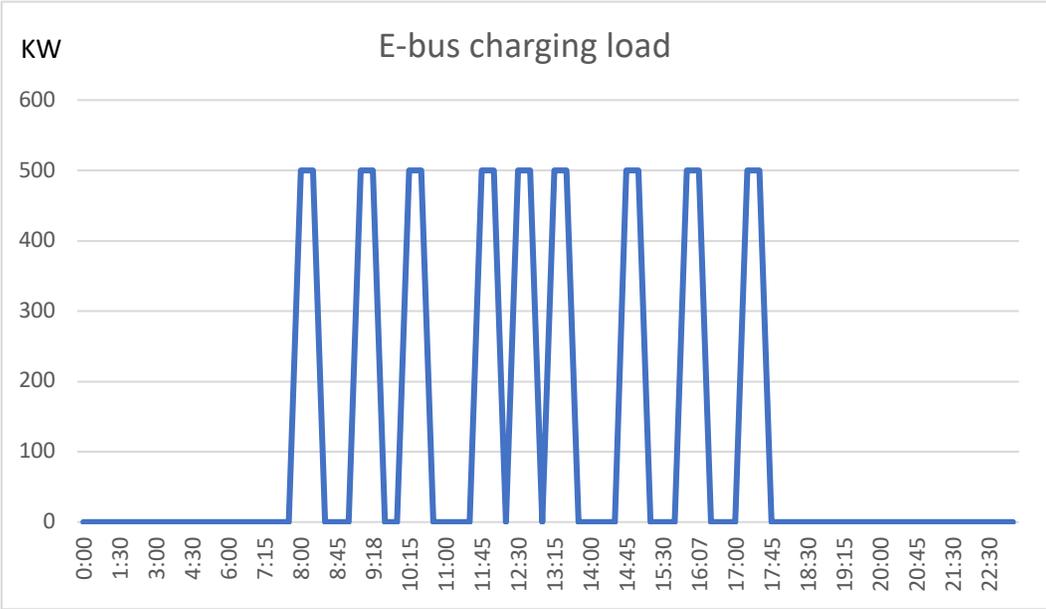

*Figure 8- E-bus charging load per kW at different times of a day.*

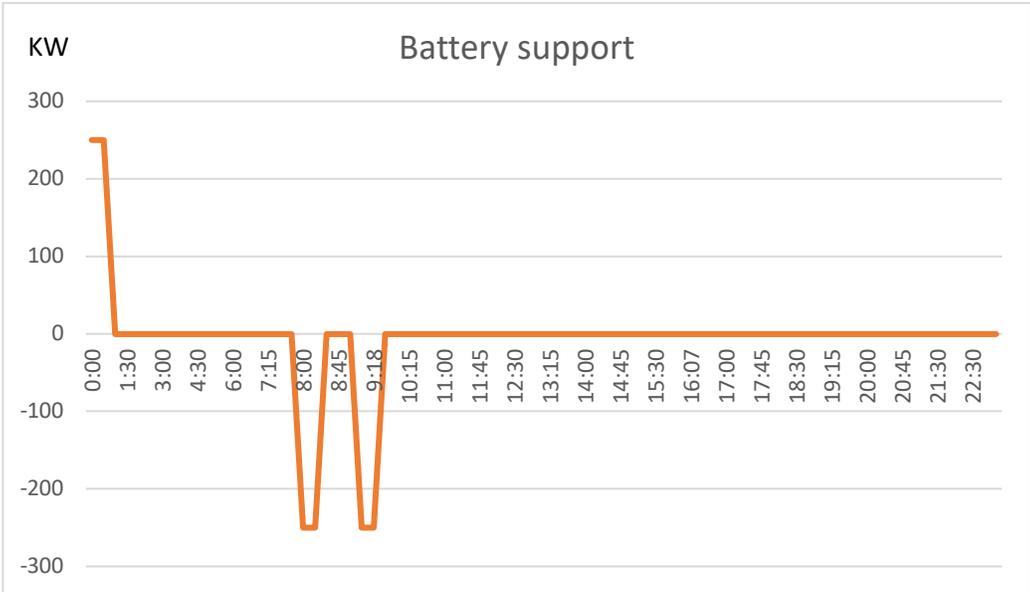

*Figure 9- Battery support per kW at different times of a day.*

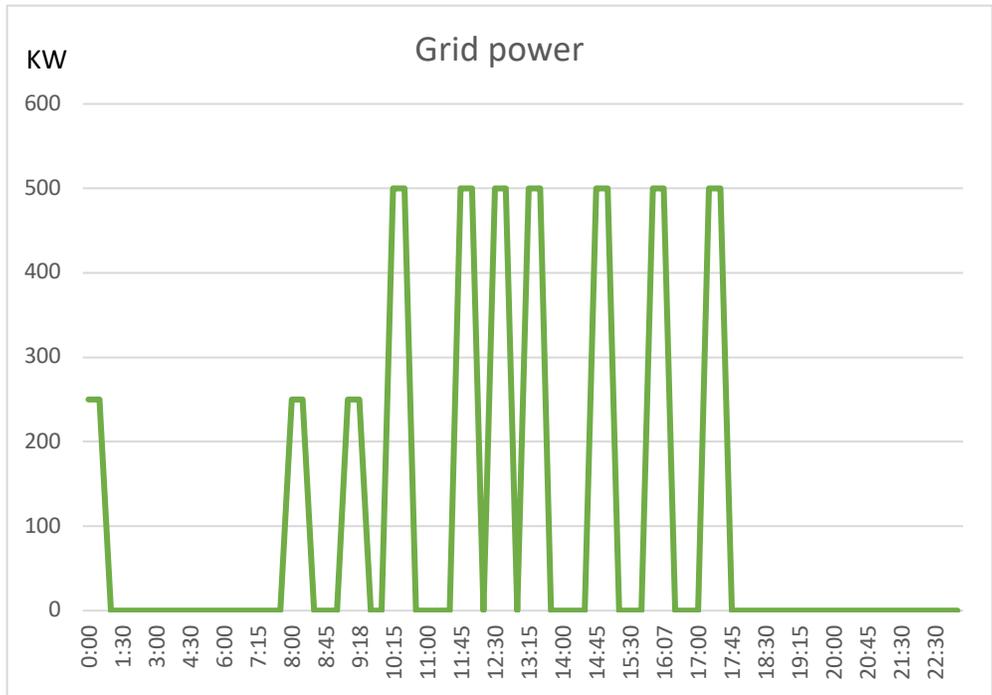
*Figure 10- Grid power per kW at different times of the day.*

The three above figures depict power load from an e-bus at different power rates. Figure 8 shows the e-bus charging load per kW at different times of the day. In order to simplify the power extracted from the grid without implementing batteries as compared to employing battery energy storage systems, the calculated power amount from the battery support can be saved in Figure 9. Also, the total power consumption from the grid perspective is drawn (Figure 10).

In the proposed software, the electricity utility cost can be entered as well. Billing is calculated for every kWh of energy used, kW of power used, and a flat rate. The input is also divided into summer and winter, and into on-peak, off-peak and mid-peak rates. The user can also define the mid-peak and on-peak hours as well as the summer months. The investment term also needs to be input for the Internal Rate of Return calculation.

Case Study and Implementation

The main focus of the project is on the electric bus charging stations, in which we are going to build up a scenario of placing 3 en-route charging stations. These charging stations will be placed in one of the routes operational by Metro buses here in LA. Also, a depot charging station will be considered somewhere close to en-route chargers to charge the buses at night. In this scenario, we will consider the typical charging time for each bus and the mileage that they will drive every day and the charging time and battery cycles.

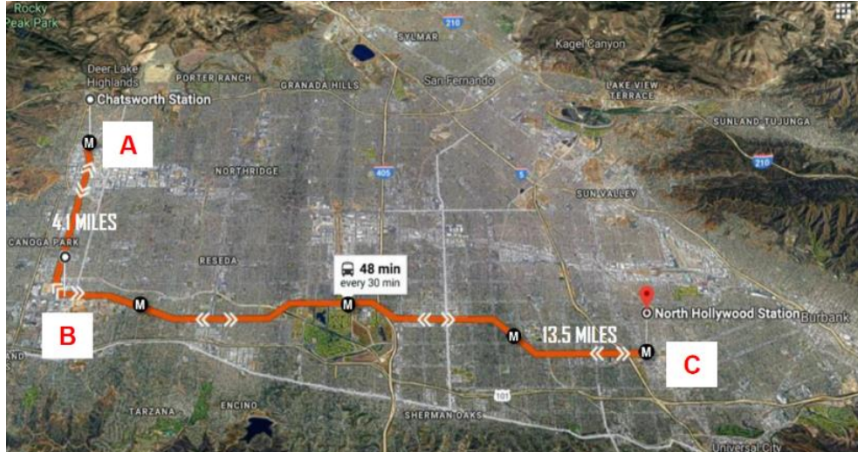
*Figure 11- The route between station A and station C [6].*

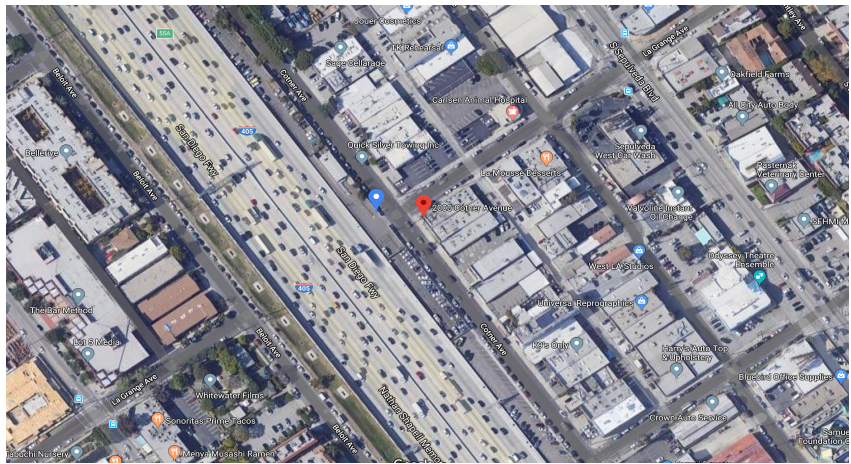
*Figure 12- Presumed pinpointed depot charging station close to i405 highway (from Google Maps).*

The presumed depot charger is located at 2000 Cotner avenue, which has easy access to the i405 highway, and there could be found plenty of parking spaces for e-bus. Depot charges are usually located where they could have access to the main route, and the land price is cheap enough to support enough spaces for buses to pull inside the station and be parked there overnight [5].

We have designed our scenario in a way that all the e-buses operate from A to C stations marked in the above figures. Also, all the e-buses will be parked at parking 8 at night, where the depot charging is, so the could be fully charged in the morning. During the day, the buses on this route can only charge via en-route charging in stations A,B, and C. The e-buses usually can be charged up within 5-10 minutes timeframe in which we calculate the pricing of electricity based on energy rates provided by SCE. The proposed scenario would attempt to minimize the charging time in a way that it won't let any two e-buses be charged at the same time which will impose a lot of load on the grid. Also, we should avoid charging during peak time as much as possible, where the electricity price would be high (peak-time). By doing so, we would be able to reduce the figures in the electricity bill for the Metro.

# Research Challenges

The long-range buses are similar in operation compared to the conventional diesel buses since their operation range is similar. This enables these buses to charge at a specific site (Depots). Different charging methods like AC slow charging or DC fast charging can be used in this case. Also, these buses are readily available, and hence their procurement and operation are simple. However, they face several disadvantages. Firstly, they cannot operate on long distance routes where it would be infeasible to use a very high capacity of the battery. Also, the batteries occupy larger space and hence reduce the passenger capacity. Furthermore, the batteries significantly increase the weight of the bus which reduces its efficiency.

The short-range buses offer better passenger capacity with reduced overall weight. However, they require intermediate high-power charging infrastructure due to their limited operating range, which requires a high initial investment.

Also, since these buses require to charge during the daytime, they cause an impact on the grid since the power usually drawn from the infrastructures is in the vicinity of 500kW [10]. When a fleet of such buses charge during the peak load time, the grid gets adversely affected and requires an additional investment of a stationary storage system to offer peak shaving.

Plus, smarter traffic designs have to be implemented to ensure proper coordination between all the vehicles and the cloud. In some cases, a change in the business model or value chain might have to be made. However, after the significant initial investment, the established charging infrastructure can be used to charge a number of buses, after which a new infrastructure will have to be developed [20, 21].

While establishing the en-route charging infrastructure, mentioned in the previous homework, a lot of factors need to be considered. Including global standardization of the charger connections and the communication protocols, feasible and optimal decisions regarding the location and the power levels for the charging, redundancies on specific routes due to an insufficient number of operating buses, and a robust charging and fleet management system to ensure proper operation of all the buses.

In general, for future developments, there are many aspects which need to be addressed [20]:

- Global standardization of the EV charger interconnection and communication protocols between charger and ebus fleet
- Business models for charging infrastructure in the multimodal transport system
- Optimal design for the best charging locations and power levels for the charging points within a predefined bus traffic system

- Dynamic considerations for redundancy and failure tolerance of the electric chargers and traffic system
- A comprehensive economic analysis and comparison study of lower TCOs of electric bus traffic system vs. diesel bus traffic system design
- ICT and cloud-based systems, connectivity with the buses, and charging infrastructure to ensure the precise scheduling of the buses.

# Future Prospect

In future work, a fleet management software will be developed to schedule the buses to follow the optimized solution by communicating using cloud systems. Further, optimizing the battery capacity with respect to cost by partial load shaving of the non-peak loads could be carried out. In order to optimize the sizing for the battery, further studies are required to understand the dynamics of electric buses. In this regard, a dynamic model of an e-bus derived from analyzing and studying the amount of e-bus electric energy consumption and stationary battery power consumption accordingly.

Stationary Battery Capacity Optimization Methodology

We have written a Matlab code for the Stationary Battery Capacity optimization. The work is still under development and will be continued in the UCLA SMERC Lab.

Direct computation of the whole problem is complicated from the understanding as well as from the debugging point of view. Hence, the main problem is split into two subproblems.

In order to reduce the load on the grid, the first subproblem is to reduce the amount of charging during peak hours. The second subproblem is to use a stationary battery to reduce or altogether remove all the power taken from the grid during peak hours. The problem needs to be solved in three steps:

Single route (End Charging points only) - single bus,
Single route (End Charging points only) - Multiple buses, and Single route (Three Charging points)
Multiple Buses and Multiple charging points (will be considered as a variable).

This problem comes under the category of Mixed Integer Linear Programming. MATLAB and Python are usually used to solve this type of problem. The parameters involved in this problem are the following:

*Table 4- Parameters used in the optimization problem.*

| Parameters | Values |
|---|---|
| Distance between the stops | 17.6 miles |
| Average **Speed** | 35.2 mph |
| Bus C/S dimensions | 3.2 m * 2.55 m (H * W) |

| Time Interval | 30 mins |
|---|---|
| Battery Capacity | 600kWh |
| Charging Rate (DC Fast) | 600kW |
| Charging Rate (Depot) | 50kW |
| Operation time | 12 hrs |

Traction Model

Traction model is used to estimate the Battery Energy Consumption. It has three components - Force due to air drag, Force due to rolling resistance, and Force due to climb resistance.

- Force due to air drag $(F_1) = 0.5 * A * C_d * \rho * v^2$
- Force due to rolling resistance $(F_2) = m * g * F_{roll} * \cos(a)$
- Force due to climb resistance $(F_3) = m * g * \sin(a)$
- Net efficiency (Motor, Power transmission and Battery) = $0.97^3 = 0.912$
- Elevation at A = 705m
- Elevation at C = 214m
- Gradient Angle (a) = atan2((705-214) /(17.6*1.6*1000)) = 1°

Apart from the above factors in energy consumption, we also need to consider the battery consumption due to the auxiliary components like the air conditioning and heater units present on the bus. We consider the power consumed by these appliances to be a constant, based on the average annual usage, and add this component to the total energy consumption after a suitable transformation of the power unit to the energy unit (kW to kWh).

Table 5- Table Input parameters.

| Mass (m) | Gradient Angle (a) | Drag Coefficient ($C_d$) | Density of air ($\rho$) | $F_{roll}$ |
|---|---|---|---|---|
| 20,000 kg | $1^0$ | 0.1 | 1.2 kg/m³ | 0.015 |

Table 6- Output parameters to calculate the Energy Consumption.

| $F_1$ | $F_2$ | $F_3$ | $F_{Net}$ | $\mathcal{E}_{Net}$ | $F_{Total}$ | Energy consumed in 30 Minutes |
|---|---|---|---|---|---|---|
| 1600 N | 2940 N | 3420 N | 7960 N | 0.92 | 8650 N | 240 KWh |

Optimization Modelling (Subproblem 1)

The Auxiliary Power (AC/Heating) is assumed to be 9 kW, which, in turn, translates to an energy of 4.5 kWh per trip.
Hence, the net energy per trip comes out to be around 245 kWh.
In this case, the objective function is to minimize the charging during peak time.
Mathematically, F = Minimize (Sum (d * 200 * Cost(t))) for 12 hrs

The constraints imposed by the system are:

1. Battery Capacity (t) = Battery Capacity (t-1) + d*300 - (1-d)*245
2. 0 <= Battery Capacity(t) <=600
3. Battery Capacity (t = 0) = 600
4. dist (t) = dist (t-1) + (1-d) * 17.6
5. dist (t=tend) >= 352 (At least 20 trips)

Here, d is the binary decision variable, which can take a value of either 0 or 1.

The first constraint ensures that the battery capacity of the bus either decreases or increases based on the binary decision variable, d. The second constraint restricts the battery capacity between the minimum and maximum value. The values usually differ from 0 and maximum battery capacity due to the depth of discharge, but this effect has been ignored over here for simplification and a better understanding of the algorithm. The third constraint shows that all the buses are fully charged when they start to operate. The fourth and fifth constraints are used to make sure that the bus travels a minimum distance and hence does the desired number of trips.

Subproblem 2

Once the optimization for the subproblem 1 is performed, the stationary battery capacity can be calculated by calculating the total energy consumed during the peak hours.

In the end, the future cost optimization problem can be summarized by the following parameters:

- Travel distance to the charging stations
- Maximum driving range for a fully-charged bus
- The initial bus range at the depot
- Recharging rate
- Driving distance between the start/endpoint of trip

- Cost of unit waiting time
- Extended driving distance
- Cost of unit driving distance
- Fixed cost per recharging activity
- Variable recharging costs in unit time
- Fixed costs per charger
- Fixed costs per charging station
- Maintenance cost per charger
- Maintenance cost per charging station

Future Recommendation for Customers

This work is aimed to benefit the electricity providers like the Los Angeles Department of Water and Power (LADWP) or Southern California Edison (SCE). As shown in the techno-economic analysis section, the main focus of this work is not to provide a cost benefit to the electricity providers but to allow them to manage the peak load demand by partially or fully supporting the e-bus charging during the peak hours. By using a Battery Energy Storage System of an optimized Battery Capacity, the electricity providers can avoid the increase in the peak load demand at the expense of some monetary loss since it is not possible to achieve both high-level performance and profit in this case [22-27].

# Conclusion

This work addresses the main issue concerning the increase in the peak load demand due to the increasing number of electric buses in the transportation systems. This research explains the different types of charging methods that are utilized in e-bus procedures. There are basically two different issues attributed to this project: 1) large battery charged with low power during night-time and/or midday breaks for a long-range e-bus 2) small battery size charged with high power rate for a short-range and city buses. It is vital to understand the critical characteristics of these two design types because of the cost structures, operation concepts in traffic system design changes accordingly.

In this research, a few parameters like the operating range, schedules, flexibility, charging time, cost, grid capacity, fleet size, charging interfaces, and passengers have been considered for the software implementation. Then we have explained the station characteristics and the demand profile that we expect in our case. The data flow analysis is explained in detail, where we have also covered the control system used in the Battery Energy Storage System and its associated Communication and Software requirements.

The techno-economic analysis found out that the total electricity bill comes out to be around $**106 per day** or an average **monthly bill of $3200**. It was found out that if the profit/cost benefit is to be realized, the battery installation should be less than $40,625. If the installation costs are higher than this, then break-even may not be brought to fruition. Then cost optimization was performed for a given charging schedule and the parameters and the algorithm were explained in detail. The output of the optimization was also explained.

As a future prospect, we have also designed an optimization methodology to optimize the battery capacity. In this case, the main problem was split into simpler subproblems, and a case scenario was assumed. Traction model was used to calculate the energy consumed by the battery per trip of the bus. Then, the objective function was formulated and the associated constraints on the system were explained.

# Acknowledgment

The authors would like to thank Dr. Rajit Gadh and Dr. Peter Chu for their kind support throughout the research.